\documentclass[twocolumn,superscriptaddress,floatfix,preprintnumbers,prl ,nofootinbib,hyperref]{revtex4-2}
\usepackage[colorlinks=true,breaklinks=true]{hyperref}
\usepackage[utf8]{inputenc}
\hypersetup{allcolors=[rgb]{0.0 0.0 0.6},linkcolor=[rgb]{0.75 0.05 0.05}}
\usepackage{mathtools}
\usepackage[dvipsnames]{xcolor}
\usepackage{float}
\usepackage{siunitx}
\usepackage{letltxmacro}
\LetLtxMacro{\oldcite}{\cite}
\renewcommand{\cite}[1]{\mbox{\oldcite{#1}}}

\DeclareSIUnit\electronvolt{e\kern-.05em V}
\DeclareSIUnit\parsec{\text{pc}}
\DeclareSIUnit\clight{\text{\ensuremath{c}}}

\usepackage{orcidlink}

\long\def\exclude#1{}

\DeclareMathOperator{\eV}{eV}

\newcommand{\beq}{\begin{equation}}
\newcommand{\eeq}{\end{equation}}

\def\ga{\,\,\raise0.14em\hbox{$>$}\kern-0.76em\lower0.28em\hbox
{$\sim$}\,\,}

\newcommand{\dd}{\mathrm{d}}
\newcommand{\mAp}{m_{A^\prime}}

\allowdisplaybreaks

\bibpunct{[}{]}{,}{n}{}{,}

\begin{document}

\title{Hints of dark photon dark matter from observations and hydrodynamical simulations of the low-redshift Lyman-$\alpha$ forest}

\author{James S. Bolton\,\orcidlink{0000-0003-2764-8248}}
\email{james.bolton@nottingham.ac.uk}
\affiliation{School of Physics and Astronomy, University of Nottingham, University Park, Nottingham, NG7 2RD, UK}

\author{Andrea Caputo\,\orcidlink{0000-0003-1122-6606}} \email{andreacaputo@mail.tau.ac.il}
\affiliation{School of Physics and Astronomy, Tel-Aviv University, Tel-Aviv 69978, Israel}
\affiliation{Department of Particle Physics and Astrophysics, Weizmann Institute of Science, Rehovot 7610001, Israel}

\author{Hongwan Liu\,\orcidlink{0000-0003-2486-0681}}
\email{hongwanl@princeton.edu}
\affiliation{Center for Cosmology and Particle Physics, Department of Physics,
New York University, New York, NY 10003, USA}
\affiliation{Department of Physics, Princeton University, Princeton, New Jersey, 08544, USA}

\author{Matteo Viel\,\orcidlink{0000-0002-2642-5707}}
\email{viel@sissa.it}
\affiliation{SISSA - International School for Advanced Studies, Via Bonomea 265, I-34136 Trieste, Italy}
\affiliation{IFPU, Institute for Fundamental Physics of the Universe, Via Beirut 2, I-34151 Trieste, Italy}
\affiliation{INFN, Sezione di Trieste, Via Valerio 2, I-34127 Trieste, Italy}
\affiliation{INAF - Osservatorio Astronomico di Trieste, Via G. B. Tiepolo 11, I-34143 Trieste, Italy}

\begin{abstract}

Recent work has suggested that an additional $\lesssim 6.9\rm\,eV$ per baryon of heating in the intergalactic medium is needed to reconcile hydrodynamical simulations with Lyman-$\alpha$ forest absorption line widths at redshift $z\simeq 0.1$. Resonant conversion of dark photon dark matter into low frequency photons is a viable source of such heating. We perform the first hydrodynamical simulations including dark photon heating and show that dark photons with mass $m_{ A'}\sim 8\times 10^{-14}\rm\,eV\,c^{-2}$ and kinetic mixing $\epsilon \sim 5\times 10^{-15}$ can alleviate the heating excess.  A prediction of this model is a non-standard thermal history for underdense gas at $z \gtrsim 3$.

\end{abstract}

\maketitle

\textit{Introduction.}---The Lyman-$\alpha$ forest, a series of absorption features that arise from the distribution of intergalactic gas, is a powerful tool for investigating the properties of dark matter (DM).  The absorption is typically observed in the spectra of distant, luminous quasars, where resonant scattering along the line of sight occurs as photons redshift into the rest frame Ly-$\alpha$ ($n = 1 \rightarrow 2$) transition of intervening neutral hydrogen \cite{McQuinn2016_review}. The 1D power spectrum of the Ly-$\alpha$ forest transmitted flux is an excellent tracer of the underlying DM distribution on scales $\sim 0.5$--$50$ comoving Mpc, and has been routinely used to place tight constraints on warm dark matter~\cite{Viel:2005qj, Boyarsky:2008xj, Viel:2013fqw, Palanque2020, Narayanan:2000tp}, fuzzy DM~\cite{Irsic:2017yje, Kobayashi:2017jcf, Armengaud:2017nkf, Rogers2021}, as well as primordial black hole (PBH) DM~\cite{Murgia:2019duy}.   

In addition, the Ly-$\alpha$ forest is also a calorimeter; the widths of the absorption lines are sensitive to the temperature of intergalactic hydrogen.  Canonically, it is assumed this temperature is set by photoelectric heating by the integrated ultraviolet (UV) emission from stars and quasars \cite{MiraldaEscudeRees1994,Puchwein2019}, and indeed, this paradigm provides an excellent match to the observed properties of the Ly-$\alpha$ forest at redshift $z>2$ \cite{Bolton2017,Villasenor2022}.  Nevertheless, the thermal state of the intergalactic medium (IGM) can also be used to constrain a variety of other possibilities, such as heating by decaying and annihilating DM~\cite{Cirelli:2009bb,Diamanti:2013bia,Liu:2016cnk,Evoli:2014pva,Lopez-Honorez:2016sur,DAmico:2018sxd,Liu:2018uzy,Cheung:2018vww,Mitridate:2018iag,Clark:2018ghm,Liu:2020wqz,haridasu20}, light PBHs~\cite{Clark:2018ghm}, DM-baryon interactions~\cite{Munoz:2017qpy}, and ultralight dark photon DM~\cite{McDermott:2019lch,Caputo:2020bdy,Caputo:2020rnx,Witte:2020rvb}.  

The physics behind Ly-$\alpha$ forest temperature measurements is straightforward. The hydrogen atoms in the IGM will in general not be at rest, but will undergo thermal motion described by a Maxwellian velocity distribution, leading to a line width $\Delta \nu = \nu_{\alpha}(b_{\rm th}^{2}+b_{\rm nth}^{2})^{1/2}/c$, where $\nu_{\alpha}$ is the resonance line frequency, $b_{\rm th} =(2 k_{\rm B} T / m_{\rm H})^{1/2} $ is the Doppler parameter due to thermal motion, $m_{\rm H}$ is the hydrogen atom mass and $k_{\rm B}$ is Boltzmann's constant.  Here, $b_{\rm nth}$ accounts for any additional, non-thermal line broadening, including, smoothing of the absorbing structures by gas pressure, small-scale turbulence, peculiar motion or expansion with the Hubble flow. $b_{\rm nth}$ is in general non-zero for all but the narrowest Ly-$\alpha$ lines. Hence, given a forward model for $b_{\rm nth}$, typically provided by cosmological hydrodynamical simulations, a determination of the Ly-$\alpha$ spectral width -- either by directly measuring Doppler parameters or using another statistic that is sensitive to the thermal broadening kernel -- allows a measurement of the IGM temperature \cite{Schaye2000,Ricotti2000,Lidz2010,Becker2011,Boera2014,Hiss2018,Walther2019,Telikova2019,Gaikwad2021}.  

Thus far, all the calorimetric constraints on new physics from the Ly-$\alpha$ forest have relied on observations at redshifts $ z \gtrsim 2$.  In this \textit{Letter}, we pioneer the use of low-redshift ($z\simeq 0.1$) Ly-$\alpha$ forest observations for this purpose.  Three independent studies \cite{Viel2017,Gaikwad2017,Burkhart:2022ygp} have now highlighted a discrepancy between the widths of Ly-$\alpha$ forest absorption lines measured from Hubble Space Telescope/Cosmic Origins Spectrograph (COS) data at $z\simeq 0.1$ \cite{Danforth2016,Kim2021} and the predictions from detailed cosmological hydrodynamical simulations.  The simulated line widths are always too narrow compared to the observations. Appealing to enhanced photoelectric heating alone is not a viable solution, as the integrated UV background would require an unphysically hard spectrum \cite{Bolton2022}.  This implies there is a non-canonical heating process in the IGM neglected in the simulations, such that an additional $\lesssim 6.9 \rm\,eV$ per baryon is deposited into typical Ly-$\alpha$ forest absorbers by $z=0.1$, and/or the non-thermal line broadening has been underestimated \cite{Bolton2022}.  Additional turbulence below the simulation resolution limit is possible \cite{Evoli2011,Iapichino2013,Burkhart:2022ygp}, although current models have line widths that become narrower, rather than broader, with increasing resolution.  Here, we instead explore the role of additional heating, and suggest that ultralight dark photon DM with a small mixing with the Standard Model (SM) photon provides an intriguing solution to the line width discrepancy.  Dark photons can undergo resonant conversions into SM photons, which are subsequently absorbed by the IGM. The condition for resonant conversion is set by the dark photon mass and the local electron density, allowing dark photons to naturally explain the low-redshift Ly-$\alpha$ forest data without altering the established agreement with IGM temperature measurements at $z>2$. 

\textit{Dark photon dark matter.}--- The dark photon $A'$ is a minimal and well-motivated extension of the Standard Model (SM) which kinetically mixes with the ordinary photon, $\gamma$~\cite{Holdom:1985ag,Dienes:1996zr,Goodsell:2009xc,Goodsell:2010ie,Goodsell:2009pi,Abel:2003ue,Abel:2006qt,Abel:2008ai}.  Ultralight dark photons are also an attractive cold DM candidate, with several early-universe production mechanisms that have been studied extensively in the literature that are capable of producing $A'$ non-relativistically~\cite{Redondo:2008ec,Nelson:2011sf,Arias:2012az,Fradette:2014sza,An:2014twa,Graham:2015rva,Agrawal:2018vin,Dror:2018pdh,Co:2018lka,Bastero-Gil:2018uel,Long:2019lwl, Caputo:2021eaa}; its perturbations are therefore expected to be well-described by the standard $\Lambda$CDM matter power spectrum. The photon-dark photon Lagrangian reads 
\begin{equation}
\mathcal{L}_{\gamma A'}=-\frac{1}{4}F_{\mu \nu}^{2}-\frac{1}{4}(F_{\mu \nu}^{\prime})^{2}-\frac{\epsilon}{2} F^{\mu \nu} F_{\mu \nu}^{\prime}+\frac{1}{2} \mAp^{2}(A_{\mu}^{\prime})^{2} \, ,
\end{equation}
where  $\epsilon$ is the dimensionless kinetic mixing parameter, and $m_{A'}$ is the $A'$ mass. $F$ and $F'$ are the field strength tensors for $\gamma$ and $A'$ respectively.  

The presence of kinetic mixing, and the resulting mismatch between the interaction and propagation eigenstates, induce oscillations of dark photons into photons, $A' \rightarrow \gamma$, and vice-versa. In the presence of a plasma, ordinary photons acquire an effective mass, $m_\gamma$, given primarily by the plasma frequency of the medium~\cite{Mirizzi:2009iz,Kunze:2015noa,Caputo:2020rnx}. At every point in space $\vec{x}$ and redshift $z$, the effective plasma mass is given by
\begin{equation}
    m_{\gamma}^2(z, \vec{x}) \simeq  \SI{1.4e-21}{\eV\squared \clight\tothe{-4}} \left(\frac{n_\mathrm{e}(z, \vec{x})}{\SI{}{\per\centi\meter\cubed}}\right) \,, \label{eq:m_gamma_sq}
\end{equation}
where $n_e$ is the free-electron number density.  

At points in space where $m_{\gamma}^2(z, \vec{x}) = \mAp^2$, the probability of conversion is resonantly enhanced~\cite{Mirizzi:2009iz,Kunze:2015noa,Caputo:2020bdy,Caputo:2020rnx}. This process can be understood as a two-level quantum system with an energy difference that is initially well-separated, but with one state having its energy evolve with time. Whenever the energy difference passes through zero, a nonadiabatic transition between the two states occurs, with transition probability described by the Landau-Zener formula~\cite{Parke:1986jy,Kuo:1989qe,Mirizzi:2009iz,Caputo:2020rnx}.   

If $A'$ constitutes the DM, the probability of conversion of $A'$ into photons is~\cite{McDermott:2019lch,Caputo:2020bdy,Caputo:2020rnx}
\begin{alignat}{1}
    P_{A' \to \gamma}(\vec{x}, t_\text{res}) \simeq \pi \epsilon^2 \frac{m_{A'} c^2}{\hbar} \left|\frac{\dd\ln m_{\gamma}^{2}(\vec{x}, t)}{\dd t}\right|_{t=t_\text{res}}^{-1}\!\!\!,
    \label{eq:prob_dm}
\end{alignat}
where $t_\text{res}$ is the time at which the resonant condition is met. Here, $h_\text{P} \equiv 2 \pi \hbar$ is Planck's constant. 

For $m_{A'}$ between \SIrange{e-15}{e-12}{\eV\per\clight\squared}, $A'$ DM converts into low-frequency photons with frequency $\nu = m_{A'} c^2 / h_\text{P}$, which rapidly undergo free-free absorption in the ionized IGM. The mean free path, $\lambda_\text{ff}$, is given approximately by~\cite{Draine2011,Chluba:2015hma}
\begin{alignat}{1}
    \lambda_\text{ff}^{-1} \simeq \frac{\alpha \sigma_\text{T} n_\text{e}^2}{2 \pi \sqrt{6 \pi}} \left( \frac{h_\text{P}}{m_\text{e} c} \right)^3 \left( \frac{k_\text{B} T}{h_\text{P} \nu} \right)^2 \left( \frac{m_\text{e} c^2}{k_\text{B} T} \right)^{7/2} \!\!\!\!\! g_\text{ff}(\nu, T) \,,
\end{alignat}
where $\alpha \simeq 1/137$ is the fine-structure constant, $\sigma_\text{T}$ is the Thomson cross section,  $T$ is the temperature of the IGM, and $g_\text{ff}(\nu, T)$ is the Gaunt factor, which only has a mild dependence on $\nu$ and $T$. Numerically,  taking $g_\text{ff} = 15.7$~\cite{2010MNRAS.406..863L}, we find
\begin{alignat}{1}
    \lambda_\text{ff} \simeq \frac{\SI{14}{\kilo\parsec}}{(1+z)^6} \, \Delta_\text{b}^{-2} \left( \frac{T}{\SI{e4}{\kelvin}} \right)^{3/2} \left( \frac{m_{A'}}{\SI{e-13}{\eV\per\clight\squared}} \right)^2 \,,
\end{alignat}
in proper units, where $\Delta_\text{b} \equiv \rho_\text{b} / \langle \rho_\text{b} \rangle$ is the local overdensity of baryons at which $A'$ conversion occurs. We assume that overdensities in both free electrons and baryons are equal, which is true for the IGM after reionization is complete. Since $\lambda_\text{ff}$ is much smaller than the typical size of a Ly-$\alpha$ forest absorber ($\sim 100 \rm\,kpc$), we can safely assume that the absorption of these photons occurs locally in our simulations. 

Since $A' \to \gamma$ conversions only occur when the resonance condition is met, at each point in time, heating only takes place in regions of specific $\Delta_\text{b}$, such that $m_{\gamma}^2(z, \vec{x}) = \mAp^2$ is satisfied. Moreover, the probability of conversion is governed by the rate at which $\Delta_\text{b}$ is evolving in each of those regions. In regions where conversions are happening, the energy deposited per baryon $E_{A' \to \gamma}$ due to such a resonance is simply $E_{A' \to \gamma} = (\rho_{A'} c^2 / n_\text{b}) P_{A' \to \gamma}$, where $n_\text{b}$ is the local number density of baryons, and $\rho_{A'}$ is the mass density of $A'$ DM. 

We can estimate $E_{A' \to \gamma}$ by assuming that the number density of baryons everywhere evolves only through adiabatic expansion, i.e.\ $n_\text{e} \propto (1+z)^3 \Delta_\text{b}$. Under this simplification, $|\mathrm{d} \ln m_\gamma^2 / \mathrm{d} t| = 3 H(z)$, where $H(z)$ is the Hubble parameter, and the probability of conversion simplifies to $P_{A' \to \gamma} = \pi \epsilon^2 m_{A'} c^2 / (3 H(z_\text{res}) \hbar)$, with $z_\text{res}$ indicating the redshift at which the resonance condition is met. For $m_{A'} \sim \SI{8e-14}{\eV}\,c^{-2}$, we expect baryons at the mean cosmological density to experience resonant conversion at $z \sim 2$. Further assuming matter domination and $\Delta_\text{dm} = \Delta_\text{b}$, we obtain approximately 
\begin{alignat}{1}
    E_{A' \to \gamma} \sim \SI{2.5}{\eV} \left(\frac{\epsilon_{-14}}{0.5}\right)^2 \left(\frac{3}{1+z_{\rm res}}\right)^{3/2} \left(\frac{m_{-13}}{0.8}\right) \,,
\end{alignat}
where $\epsilon_{-14} \equiv \epsilon / 10^{-14}$, and $m_{-13} \equiv m_{A'} / (\SI{e-13}{\eV}\,c^{-2})$. We see that even a mixing as small as $\epsilon_{-14} \sim 1$ leads to the absorption of enough photons to heat the IGM by several \SI{}{\eV} per baryon~\cite{McDermott:2019lch,Caputo:2020rnx,Caputo:2020bdy, Witte:2020rvb}.

The resonant nature of $A'$ heating makes it an attractive model for reconciling low and high redshift Ly-$\alpha$ forest data. The smaller the value of $m_{A'}$, the later the resonance condition is met at fixed $\Delta_\text{b}$. For a sufficiently light $A'$, most of the resonant conversions occur at $z \lesssim 2$, broadening the absorption line widths at low redshifts, without depositing a significant amount of heat at $z > 2$. Furthermore, the resonance condition is set by $n_e \propto (1+z)^3 \Delta_\text{b}$, implying that $(1+z_\text{res}) \propto \Delta_\text{b}^{-1/3}$ and $E_{A' \to \gamma} \propto \Delta_\text{b}^{1/2}$ during matter domination, giving approximately the density dependence required by observations~\cite{Bolton2022}.

\textit{Simulations.}---To fully test the viability of heating from $A'$ dark matter, we now turn to implementing this model in Ly-$\alpha$ forest simulations for the first time. Cosmological hydrodynamical simulations of the Ly-$\alpha$ forest were performed with a version of P-Gadget-3~\cite{2005MNRAS.364.1105S}, modified
for the Sherwood simulation project~\cite{Bolton2017}. Following \cite{Bolton2022}, we use a simulation box size of $L=60h^{-1}\rm\,cMpc$ with $2\times 768^{3}$ gas and dark matter 
particles, giving a gas (dark matter) particle mass of $M_{\rm
  gas}=6.38\times 10^{6}h^{-1}M_{\odot}$ ($M_{\rm dm}=3.44\times
10^{7}h^{-1}M_{\odot}$). The simulations were started at $z = 99$, with initial conditions generated on a regular grid using a $\Lambda$CDM transfer function generated by CAMB \cite{Lewis:1999bs}.  The cosmological parameters we use are
$\Omega_{\rm m}=0.308$, $\Omega_{\Lambda}=0.692$, $h=0.678$,
$\Omega_{\rm b}=0.0482$, $\sigma_{8}=0.829$ and $n=0.961$ \cite{Planck2014}, with a primordial helium fraction by mass of
$Y_{\rm p}=0.24$. All gas particles with overdensity $\Delta_{\rm b} > 10^{3}$ and temperature $T<10^{5}\rm\,K$ are converted into collisionless star particles, and photo-ionization and heating by a spatially uniform UV background is included \cite{Puchwein2019}.  Mock Ly-$\alpha$ forest spectra were extracted from the simulations and processed to resemble COS observational data following the approach described by \cite{Bolton2022}, where further details and tests of our numerical methodology can be found.

In contrast to \cite{Bolton2022}, however, in this work we also implement dark photon heating in our hydrodynamical simulations using Eq.~(\ref{eq:prob_dm}).  It will be numerically convenient to assume the baryons closely trace the dark matter in the IGM, and indeed, on scales exceeding $\sim 100 \rm\,kpc$ (the pressure smoothing scale in the IGM), this is a good approximation. Defining the overdensity of a given cosmological species as $\Delta_i = \rho_i/ \langle \rho_i \rangle$, we thus assume $\Delta_{\rm dm} = \Delta_\text{b} = \rho_\text{b}/ \langle \rho_\text{b} \rangle$ for $\langle \rho_\text{b} \rangle = \rho_{\rm crit} \Omega_\text{b} (1+z)^3$, where $\rho_\text{b} \equiv \rho_\text{b}(\vec{x}, z)$ is determined for each gas particle at position $\vec{x}$ and redshift $z$ in our simulations.  For each gas particle, we then set $\rho_{A'}(\vec{x}, z) = \Delta_\text{b}(\vec{x}, z)\rho_{\rm crit}(\Omega_{\rm m}-\Omega_{\rm b}) (1+z)^3$.  For a fixed value of $m_{A'}$, we may therefore determine where and when a resonant conversion happens for each gas particle (i.e.\ when the condition $m_\gamma^{2}(\vec{x}, z_{\rm res}(\vec{x})) = \mAp^{2}$ is met).  The converted energy per baryon $E_{A' \to \gamma}$ is then calculated using Eq.~(\ref{eq:prob_dm}), and directly injected into the gas particles.  

\begin{figure}
\begin{center}
  \includegraphics[width=0.48\textwidth]{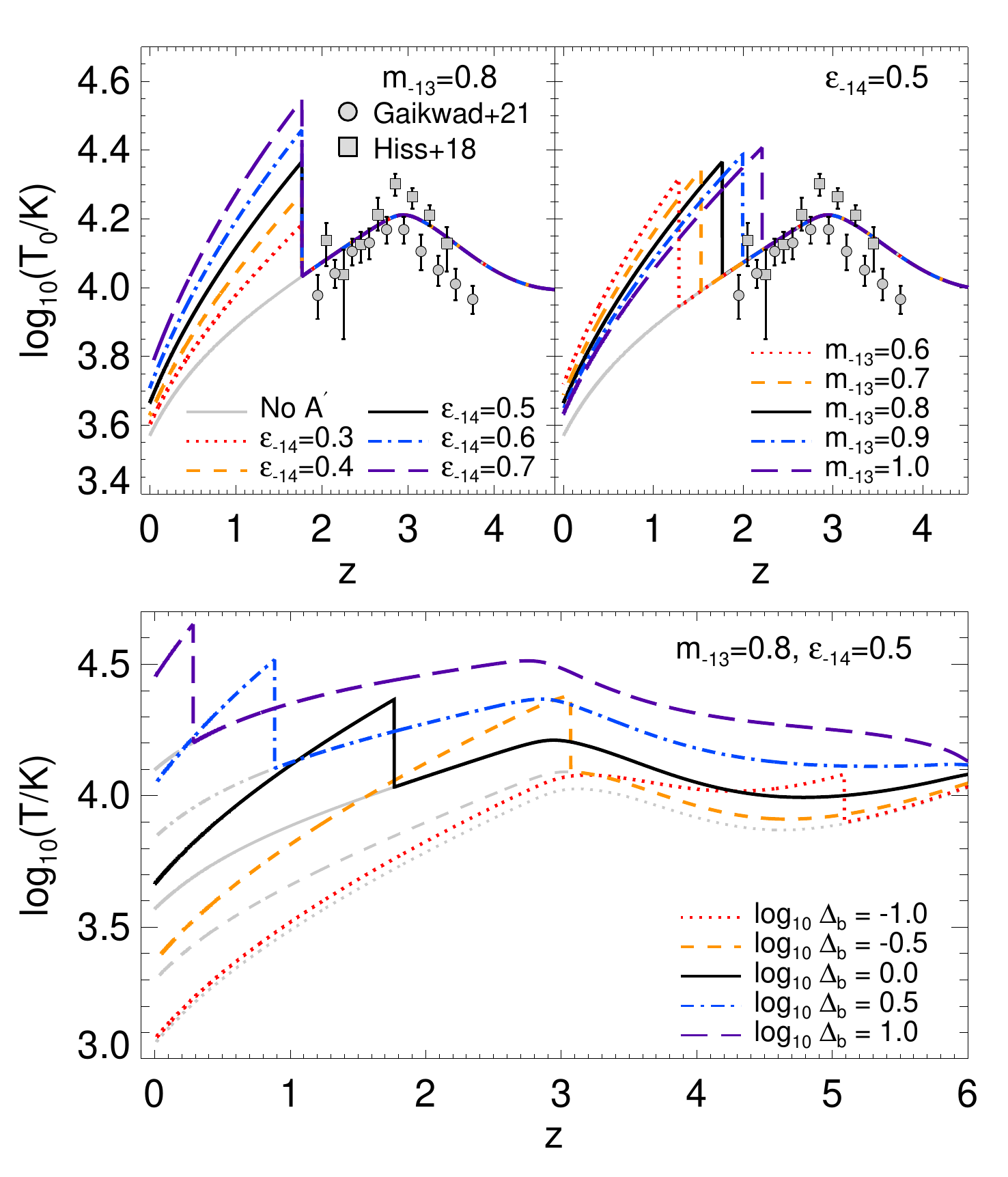}
  \vspace{-0.85cm}
  \caption{\emph{Upper panels:} Thermal histories for baryons at the mean background density, $\Delta_{\rm b}=\rho_{\rm b}/\langle \rho_{\rm b} \rangle=1$. Data points show the IGM temperature measurements from \cite{Hiss2018,Gaikwad2021}. The gray solid curves show the case for no dark photon heating, while the other curves illustrate the effect of varying $\epsilon$ (left) or $m_{\rm A^{\prime}}$ (right). 
  \emph{Lower panel:} Thermal histories for baryons at fixed overdensities, $\Delta_{\rm b}$, assuming $m_{-13}=0.8$ and $\epsilon_{-14}=0.5$. Gray curves again show the case for no dark photon heating.}
  \label{fig:tevol}
\end{center}
\end{figure}

\textit{Results.---} We first demonstrate the effect that dark photons have on the IGM temperature by considering the redshift evolution of gas parcels at fixed overdensity, $\Delta_{\rm b}$, heated by both the UV background and $E_{A' \to \gamma}$.  We adapt the non-equilibrium ionization and heating calculations performed by \cite{Bolton2022} for this purpose. In the upper panels of Fig.~\ref{fig:tevol} we show the thermal history of gas at the mean density, $\Delta_{\rm b} = 1$.  The solid gray curves correspond to UV heating only (i.e.\ no $A'$), following the synthesis model presented in \cite{Puchwein2019}.  The data points are IGM temperature measurements at the mean density derived from the Ly-$\alpha$ forest \cite{Hiss2018,Gaikwad2021}.    All other curves in Fig.~\ref{fig:tevol} include $A'$ heating.  These exhibit a sharp rise in the gas temperature when the resonance condition is met. In the top left panel, we have fixed the $A'$ mass to $m_{-13} = 0.8$ and varied the kinetic mixing parameter, $\epsilon$. In this case, the timing of the energy injection does not change, but the amplitude of the temperature peak increases with $\epsilon$. The top right panel instead shows the results for a fixed kinetic mixing, $\epsilon_{-14} = 0.5$, for different $A'$ masses.  In this case,  smaller $A'$ masses result in later injection of heat into the gas parcel. 
From this, we may already conclude that $A'$ masses with $m_{-13}\gtrsim 0.9$ and $\epsilon_{-14} = 0.5$ are excluded by the data from \cite{Hiss2018,Gaikwad2021} at $2 < z < 4$. This is consistent with the bounds derived analytically in \cite{Caputo:2020bdy} using earlier measurements of the mean density IGM temperature (see their Fig.~9, as well as \cite{McDermott:2019lch,Witte:2020rvb}). 

\begin{figure}
\begin{center}
  \includegraphics[width=0.48\textwidth]{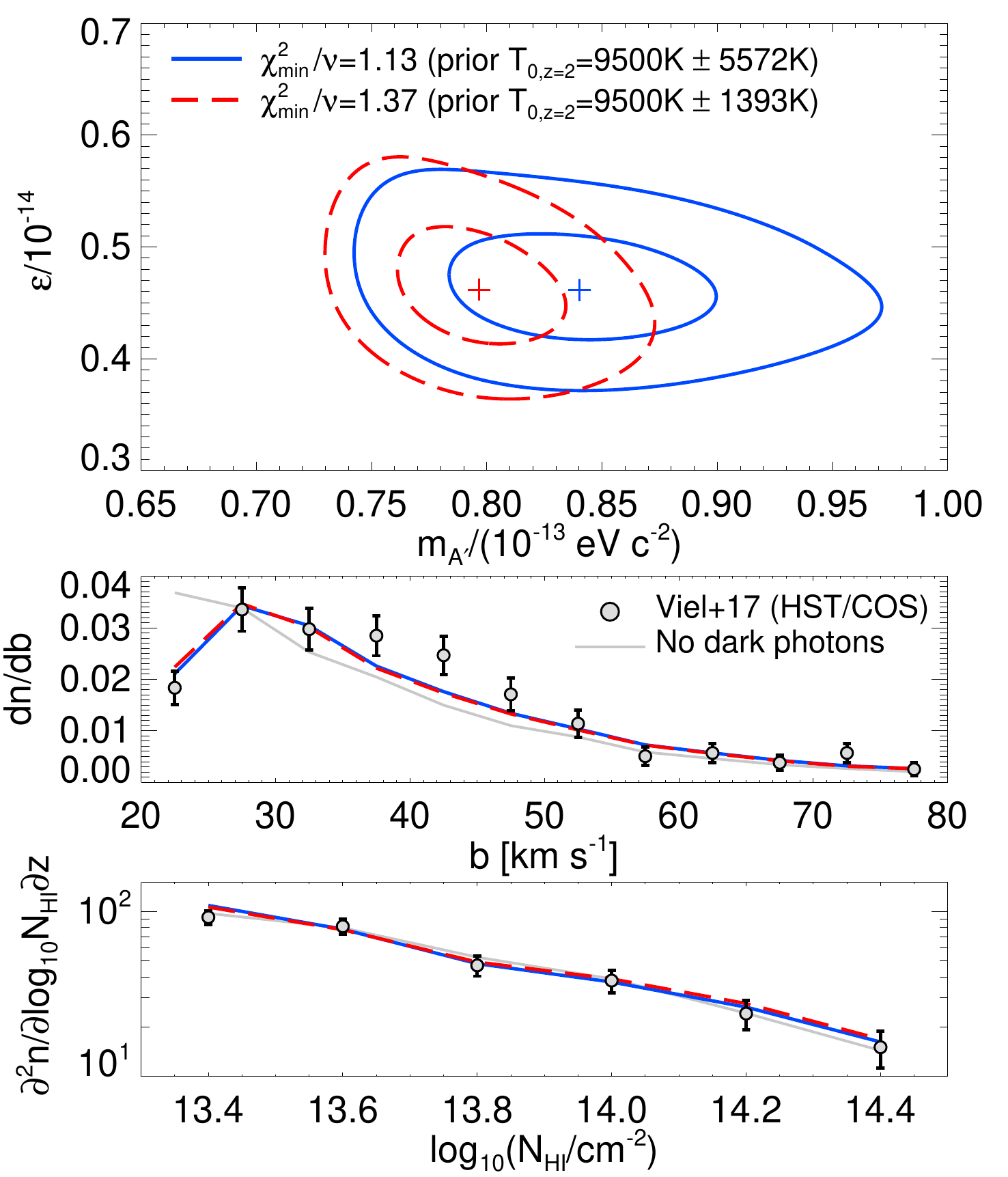}
  \vspace{-0.7cm}
  \caption{\emph{Upper panel:} Fit to the $b$-parameter distribution and CDDF of the Ly-$\alpha$ forest at $z=0.1$ \cite{Viel2017,Kim2021} assuming a \emph{maximal} contribution of dark photon heating to the line widths.  Contours show $\Delta \chi^{2}=\chi^{2}-\chi_{\rm min}^{2}=1$ and $4$, corresponding to the projection of the 68\% and 95\% intervals for the individual parameters $m_{A'}$ and $\epsilon$.  The dashed red curves show the results for a tight prior on the $z=2$ IGM temperature at mean density from ~\cite{Gaikwad2021}. 
  The solid blue curves show the effect of a weaker prior, where the 1$\sigma$ uncertainty from ~\cite{Gaikwad2021} has been increased by a factor of four for consistency with the independent temperature measurement from~\cite{Hiss2018}. 
  \emph{Lower panels:} The corresponding best-fit models compared to the COS observational data.  The solid gray curve shows the UV heating model with no dark photon heating~\cite{Puchwein2019}.}
  \label{fig:constraint}
\end{center}
\end{figure}

The lower panel of Fig.~\ref{fig:tevol} shows the gas thermal evolution for a fixed pair of parameters $[m_{-13}, \epsilon_{-14}] = [0.8, 0.5]$, but now for varying gas overdensities. For the adopted value of the $A'$ mass, $m_{-13}= 0.8$, the resonance at the mean background density occurs at $z_{\rm res}=1.8$  (black solid curve).  Later energy injection occurs for overdensities ($\log_{10} \Delta_{\rm b}>0$), while earlier energy injection occurs for underdensities ($\log_{10} \Delta_{\rm b}<0$). This density dependence allows low-redshift Ly-$\alpha$ forest observations to place a bound on $A'$ heating that is complementary to constraints obtained at $z>2$.  The Ly-$\alpha$ forest at $z=0.1$ is sensitive to gas at overdensities of $\Delta_{\rm b}\simeq 10$ \cite{Bolton2022}, whereas at $z>2$, it probes gas close to the mean density \cite{Becker2011}.  Hence, if appealing to $A'$ heating as a possible resolution to the COS line width discrepancy, we require $m_{-13} > 0.6$ for resonant conversion to occur in gas with $\Delta_{\rm b}=10$ by $z=0.1$; smaller masses would inject the energy too late.  Coupled with the upper bound from data at $2<z<4$, this implies $0.6 \lesssim m_{-13}\lesssim 0.9$ for all kinetic mixing parameters that heat the IGM by more than a few thousand degrees.   Note also that heating of the \emph{underdense} IGM is expected at $z>2$, even if the mean density IGM temperature constraints at $2<z<4$ are fully satisfied.   Intriguingly, there are indeed hints from the distribution of the Ly-$\alpha$ forest transmitted flux that underdense gas in the IGM at $z=3$ is hotter than expected in canonical UV photo-heating models \cite{Bolton2008,Rorai2017_pdf}.  We intend to explore this further in future work.

We now turn to obtaining a best-fit dark photon dark matter model from the low-redshift Ly-$\alpha$ forest assuming \emph{maximal} $A'$ heating, i.e.\ with no other sources of noncanonical heating.  Using the recipe outlined in the previous section, we have performed nine hydrodynamical simulations with different values of the $A'$ mass, $m_{-13}=[0.6,\, 0.8,\, 1.0]$, and kinetic mixing, $\epsilon_{-14}=[0.3,\,0.5,\,0.7]$.  Voigt profiles were fit to the mock Ly-$\alpha$ spectra, giving the Ly-$\alpha$ line column densities, $N_{\rm HI}$, and Doppler parameters, $b$.  The $b$-distribution and column density distribution function (CDDF) were then  constructed for each simulation following \cite{Bolton2022}, and the model grid was used to perform a $\chi^{2}$ minimisation on the covariance matrix derived from the COS observations.

The resulting best fit $A'$ parameters are presented in Fig.~\ref{fig:constraint}.  The $\Delta \chi^{2}=\chi^{2}-\chi_{\rm min}^{2}$ contours corresponding to the 68\% and 95\% intervals for individual parameters are displayed in the upper panel, while the lower panels show the best-fit models (corresponding to the crosses in the upper panel).  We assume two different priors on the temperature of the IGM at $z=2$: a tight prior of $T_{0,z=2}=9500\pm \SI{1393}{\kelvin}$ based on \cite{Gaikwad2021} (red dashed contours), and a weaker prior where the $1\sigma$ uncertainty on $T_{0,z=2}$ has been increased by a factor of four for consistency with the independent measurement of $T_{0,z=2}=13721^{+1694}_{-2152} \, \SI{}{\kelvin}$ from \cite{Hiss2018} (blue solid contours).  As expected from Fig.~\ref{fig:tevol}, a weaker prior has the effect of increasing the best-fit $A'$ mass.  For the weak (tight) $z=2$ prior, the best-fit model has $m_{-13}=0.84\pm0.06$ ($m_{-13}=0.80\pm0.04$) and  $\epsilon_{-14}=0.46_{-0.04}^{+0.05}$ ($\epsilon_{-14}=0.46_{-0.05}^{+0.06}$)  ($1\sigma$) for $\chi^{2}_{\rm min}/\nu=1.13$ ($\chi^{2}_{\rm min}/\nu=1.37$) and $\nu=16$ degrees of freedom, with a $p$-value of $p=0.32$ ($p=0.14)$.  For comparison, a model with UV photon heating only \cite{Puchwein2019}, shown by the gray curves in the lower panels of Fig.~\ref{fig:constraint}, has $\chi^{2}_{\rm min}/\nu=3.50$ and $p=2.5\times 10^{-6}$ assuming the weak $z=2$ thermal prior.  The addition of $A'$ heating considerably improves the fit and leads to very good agreement with the COS data.  For the weak $z=2$ prior, the best-fit $A'$ parameters deposit an extra \SI{5.3}{\eV} per baryon into gas with $\Delta_{\rm b}=10$ by $z=0.1$, consistent with the limit of $\lesssim \SI{6.9}{\eV}$ per baryon obtained by \cite{Bolton2022}.

\textit{Conclusions.---} In this \textit{Letter} we have pioneered the use of the Ly-$\alpha$ forest at redshift $z\simeq 0.1$ as a calorimeter for investigating properties of the dark sector. Specifically, we studied a model of ultralight dark photons, $A'$, that can naturally alleviate the tension \cite{Viel2017,Gaikwad2017,Bolton2022,Burkhart:2022ygp} between the (too narrow) Ly-$\alpha$ absorption line widths predicted in hydrodynamical simulations compared to observational data at $z=0.1$.   Assuming a \emph{maximal} contribution from $A'$ heating and a thermal prior of $T_{0}=9500 \pm 5572\rm\,K$ at $z=2$ \cite{Gaikwad2021,Hiss2018}, our best-fit model has $A'$ mass $m_{A'}=8.4\pm0.6\times 10^{-14}\rm\,eV\,c^{-2}$ and kinetic mixing parameter $\epsilon=4.6^{+0.5}_{-0.4}\times 10^{-15}$ ($1\sigma$).   Although astrophysical sources, such as turbulent broadening or other non-canonical heating processes may also explain the line width discrepancy, our study is a first clear indication that DM energy injection can be a compelling alternative.  We also highlight that our best-fit $A'$ parameters will have testable consequences for the temperature of the underdense IGM at $z=3$, where there are already hints of missing heating in Ly-$\alpha$ forest simulations \cite{Bolton2008,Rorai2017_pdf}.  

Finally, we note that dark photons in our mass range of interest can be produced around spinning black holes (BHs) through the superradiance instability. The superradiant cloud can affect the spin-mass distribution of BHs or produce gravitational waves, giving a promising way to look for $A'$ with $m_{A'} \sim \SI{8e-14}{\eV}\,c^{-2}$. At the moment, BH measurements appear to be in tension with this $A'$ mass~\cite{Baryakhtar:2017ngi,Cardoso:2018tly,Ghosh:2021zuf}, but are currently subject to significant uncertainties~\cite{Belczynski:2021agb,Ghosh:2021zuf} and model dependence~\cite{Fukuda:2019ewf, Baryakhtar:2017ngi, Caputo:2021efm, Cannizzaro:2022xyw}. Sharpening our understanding of the superradiance phenomenon, as well as improving the experimental searches, will be pivotal in testing our model. 
  
\textit{Acknowledgments.---}%
AC is supported by the Foreign Postdoctoral Fellowship Program of the Israel Academy of Sciences and Humanities and also acknowledges support from the Israel Science Foundation (Grant 1302/19) and the European Research Council (ERC) under the EU Horizon 2020 Programme (ERC-CoG-2015-Proposal n. 682676 LDMThExp). HL is supported by NSF grant PHY-1915409, the DOE under Award Number DE-SC0007968 and the Simons Foundation. MV is supported by the PD51-INFN INDARK and ASI-INAF n. 2017-14-H.0 grants. JSB is supported by Science and Technology Facilities Council (STFC) consolidated grant ST/T000171/1.  The hydrodynamical simulations were performed using the DiRAC@Durham facility managed by the Institute for Computational Cosmology on behalf of the STFC DiRAC HPC Facility. The equipment was funded by BEIS capital funding via STFC capital grants ST/P002293/1 and ST/R002371/1, Durham University and STFC operations grant ST/R000832/1.




\bibliographystyle{bibi}
\bibliography{biblio}

\onecolumngrid
\appendix

\clearpage

\setcounter{equation}{0}
\setcounter{figure}{0}
\setcounter{table}{0}
\setcounter{page}{1}
\makeatletter
\renewcommand{\theequation}{S\arabic{equation}}
\renewcommand{\thefigure}{S\arabic{figure}}
\renewcommand{\thepage}{S\arabic{page}}

\begin{center}

\textbf{\large Hints of dark photon dark matter from observations and hydrodynamical simulations of the low-redshift Lyman-$\alpha$ forest}

\vspace{0.05in}

\textit{\large Supplemental Material}

\vspace{0.05in}

{James S. Bolton, Andrea Caputo, Hongwan Liu and Matteo Viel}

\end{center}


\bigskip



In this Supplemental Material, we show results from a wider range of our hydrodynamical simulations that use different values of the $A'$ mass and kinetic mixing.  In contrast to Fig.~\ref{fig:constraint} in the main text, rather than show the best fit model, we show results from the individual simulations that span a range of $A'$ masses and kinetic mixings.  As already discussed in the main text, this demonstrates the effectiveness of the low-redshift Ly-$\alpha$ forest data for obtaining best fit parameters for $A'$, rather than just individuating a preferred region of the parameter space.

\begin{figure}[H]
\begin{center}
  \includegraphics[width=0.49\textwidth]{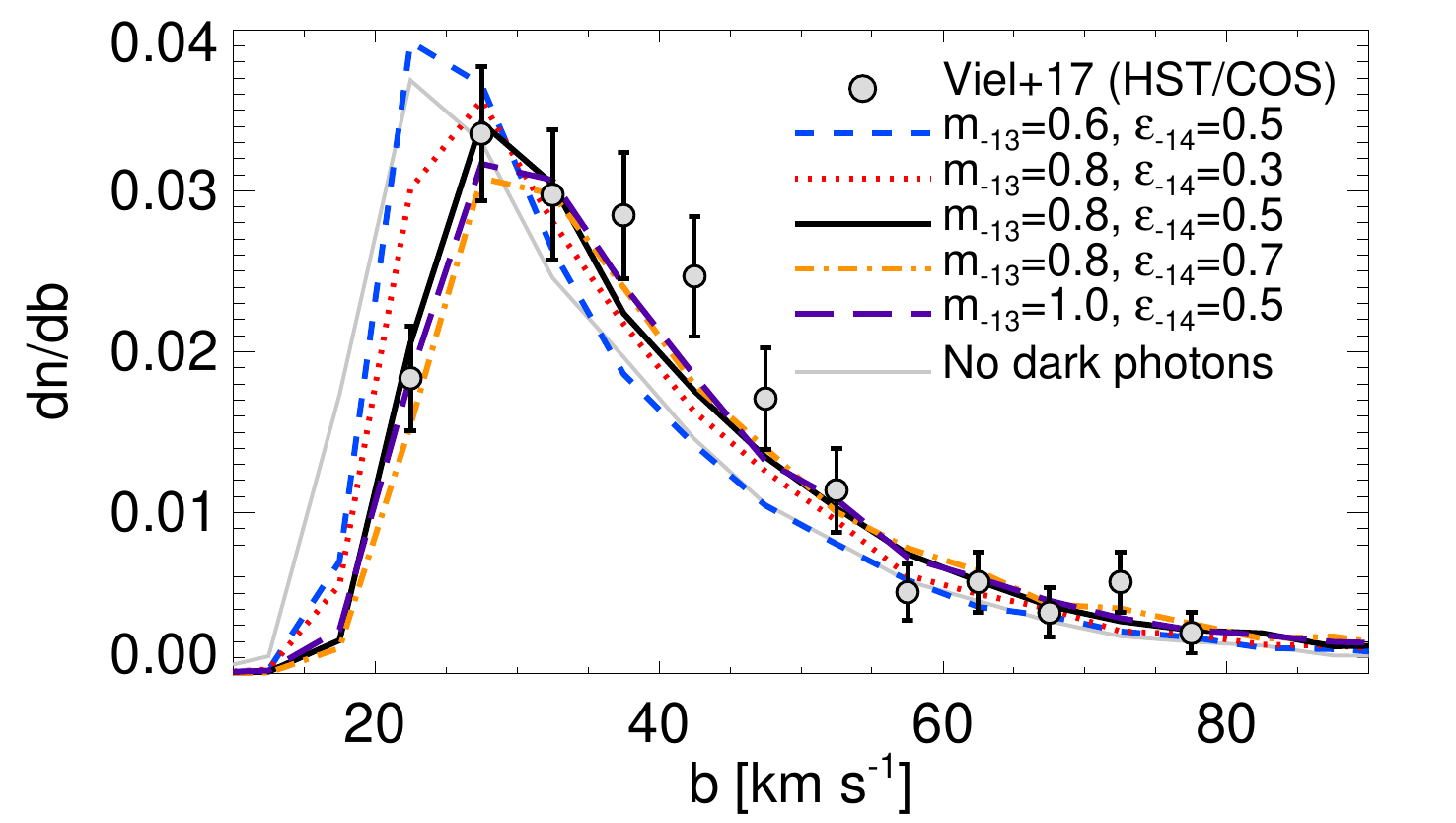}
   \includegraphics[width=0.49\textwidth]{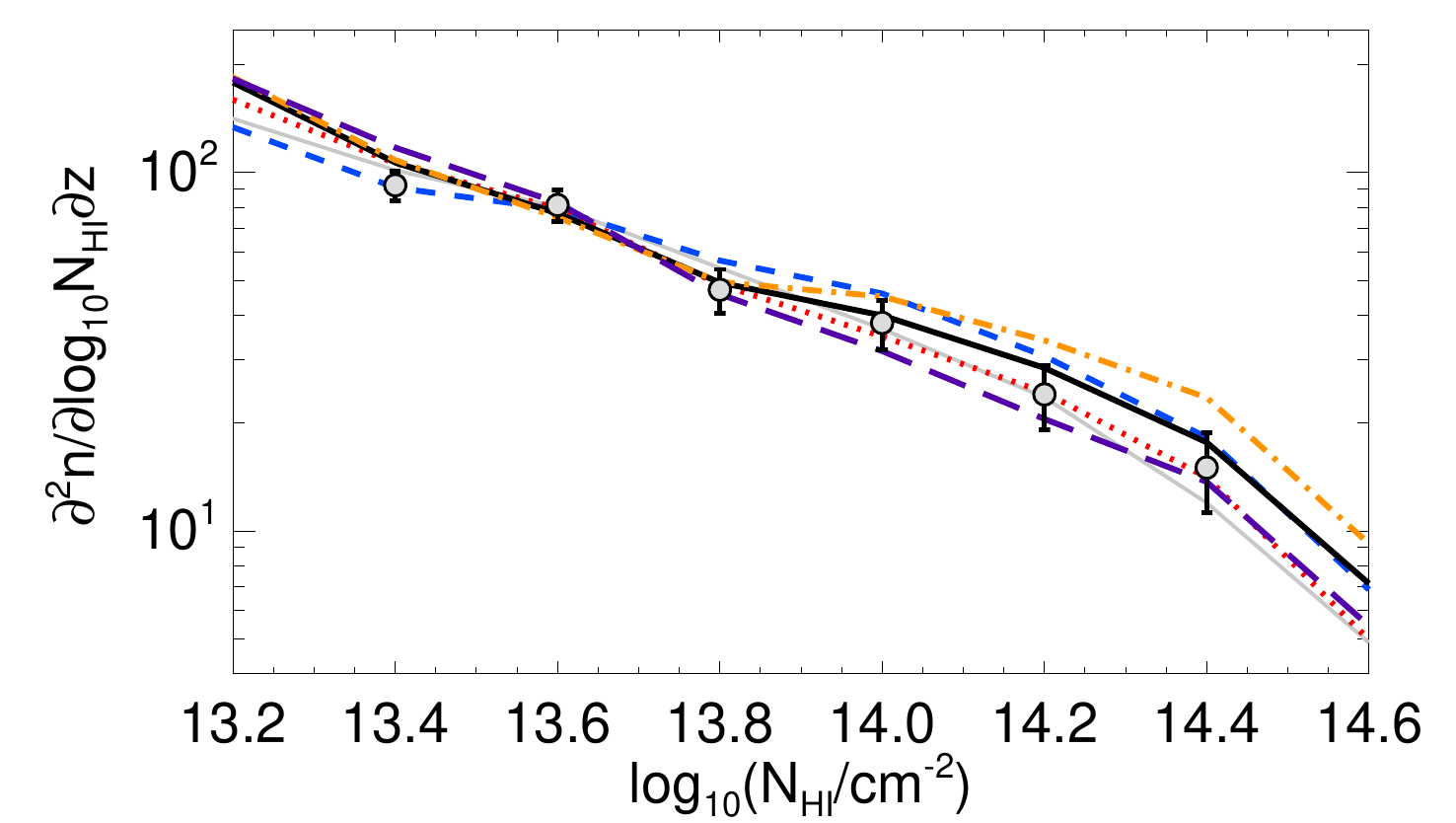}
  \vspace{-0.2cm}
  \caption{Illustration of the effect of the $A'$ mass, 
  and the kinetic mixing parameter 
  on the $z=0.1$ Ly-$\alpha$ forest $b$-parameter distribution (left panel) and CDDF (right panel) predicted by hydrodynamical simulations.   The gray data points show the COS observational measurements from \cite{Viel2017}. Solid gray curves show a simulation with UV photon heating only \cite{Puchwein2019}. Note that the dark photon heating impacts both the Ly-$\alpha$ line widths \emph{and} the shape of the CDDF, making a joint fit particularly powerful.
  }
  \label{fig:MoreResults}
\end{center}
\end{figure}

Fig.~\ref{fig:MoreResults} shows the effects of dark photon energy injection on the $z=0.1$ Ly-$\alpha$ forest $b$-parameter distribution (left panel) and column density distribution function (CDDF, right panel) predicted by the cosmological hydrodynamical simulations.  Note that, following \cite{Viel2017,Bolton2022}, the neutral hydrogen densities in the models have been linearly rescaled to approximately match the amplitude of the CDDF at $\log_{10}(N_{\rm HI}/\rm cm^{-2})\sim 13.5$ (corresponding to a baryon overdensity $\Delta_{\rm b}\sim 10$).  The only difference here is we consider Doppler parameters in the range $\SI{20}{\kilo\meter\per\second} \leq b \leq  \SI{80}{\kilo\meter\per\second}$ (cf. $\SI{20}{\kilo\meter\per\second} \leq b \leq \SI{90}{\kilo\meter\per\second}$ in \cite{Bolton2022}); the largest Doppler parameters are highly suprathermal and are insensitive to $A'$ heating. 

From the $b$-parameter distribution in the left panel of Fig.~\ref{fig:MoreResults}, it is clear that $A'$ with mass $m_{-13}=0.6$  does not give substantially better agreement with the data compared to the case without $A'$ heating (gray line); both models overshoot the data for $b<25\rm\,km\,s^{-1}$.  In this particular case, the $A'$ heating occurs too late to affect the typical gas densities probed by the Ly-$\alpha$ forest at $z=0.1$ (for $m_{-13}=0.6$ the resonance redshift is $z_{\rm res}=0.06$ for $\Delta_{\rm b}=10$).
Notice also that some of the $A'$ parameters excluded by the $b$-parameter distribution provide a reasonable fit to CDDF, while, on the other hand, some parameters that yield a reasonable fit to $b$-parameter distribution data are inconsistent with the CDDF data (see for example the orange dot-dashed curve in Fig.~\ref{fig:MoreResults}).   In the particular case of $m_{-13}=0.8$ (corresponding to $z_{\rm res}=0.28$ for $\Delta_{\rm b}=10$), gas with $\log_{10}(N_{\rm HI}/\rm cm^{-2})\sim 13.5$ ($\Delta_{\rm b}\sim 10$) is hotter in the models with larger kinetic mixing.  However, gas at $\Delta_{\rm b}\gtrsim 16$ (i.e.\ densities where $z_{\rm res}<0.1$ for $m_{-13}=0.8$) corresponding to the higher column density Ly-$\alpha$ absorbers) is not heated because the resonance threshold has not been crossed.  This has the effect of flattening the gradient of the CDDF; if the lower density gas is hotter, it is also more ionized (the case-A recombination rate for hydrogen, $\alpha_{\rm A}\propto T^{-0.72}$, where $T$ is the gas temperature).  This results in relatively more of the stronger lines with $\log_{10}(N_{\rm HI}/\rm cm^{-2})>13.8$ as the kinetic mixing is increased.  This demonstrates the interplay between these two observables which, when combined, provide a more powerful test of $A'$ heating than either measurement alone. 

\end{document}